\documentclass{iopart}
\def\text{\rm}
\usepackage{graphicx}
\begin{document}

\title{Atom loss from the $^{85}$Rb Bose-Einstein condensate by a
 Feshbach
resonance}

\author{V. A. Yurovsky and A. Ben-Reuven}

\address{School of Chemistry, Tel Aviv University, 69978 Tel Aviv,
Israel}

\begin{abstract}

Loss of atoms from a $^{85}$Rb condensate on passage through a
Feshbach resonance is analyzed using the generalized parametric
approximation that takes into account quantum many-body effects.
These effects lead to a substantial increase of the losses. A 
better agreement with experiments is achieved, compared to 
predictions of mean-field theories. The 
method provides much insight into the quantum effects involved, 
and on the nature of entangled atom pairs produced by the loss.

\end{abstract}
\submitto{\JPB}

The presence of a Feshbach resonance is supposed to provide a
tool for controlling the interatomic interactions determining the
properties of a Bose-Einstein condensate (BEC) (see \cite{TTHK99}
and references therein). A Feshbach resonance occurs when the
energy of a pair of atoms in the condensate is close to that of a
metastable molecular state. The scattering length, as a measure of
the mean interatomic interaction, varies strongly as a function of
the energy mismatch between the two states. This mismatch can be
controlled by applying a varying magnetic field. The energies of
the two states can be brought closer to each other, as the two
states have different Zeeman shifts.

The effect was studied first in the MIT experiment on Na
\cite{IASMSK98}, by applying a time-varying magnetic field $B=B\left(
 t\right) $. The
experiment resulted in a large condensate population loss. In order to
provide a theoretical explanation of these experimental results, two
mechanisms have been suggested. The first one is a collisional
deactivation process \cite{TTHK99,YBJW99}, relating the loss to 
atom-molecule and molecule-molecule inelastic collisions. This mechanism
has been analyzed in \cite{TTHK99,YBJW99} using coupled mean-field
(MF) equations for the atomic and molecular condensates. The second
mechanism is an excitation process \cite{MTJ00}, involving a crossing
of the resonant molecular state into non-condensate atomic states. It
has been analyzed at first in \cite{MTJ00} as a dissociation of single
molecules, without taking into account many-body effects. The combined
effect of both mechanisms was studied in \cite{YBJW00}, where the
crossing mechanism has been incorporated into MF equations by
introducing a width to the molecular condensate state. It has been
shown that in the case of the MIT experiment both mechanisms
contribute to the loss comparably and non-additively. The crossing
loss mechanism and non-condensate states are especially important in
the case of the extremely strong Feshbach resonance in $^{85}$Rb 
studied in the more recent JILA experiments \cite{C00}.

The non-condensate atoms are formed as entangled pairs in two-mode
squeezed states \cite{YB03}. Several methods have been suggested 
earlier, allowing a correct treatment of such essentially quantum 
states. A numerical solution of stochastic differential equations 
in the positive-$P$ representation has been used for this purpose in 
\cite{PM01}. In the Hartree-Fock-Bogoliubov formalism (HFB) 
\cite{HPW01,KH02} the MF equations were complemented by
equations for the normal and the anomalous densities, describing the
second-order correlations of the non-condensate atomic fields. These 
correlations are taken into account to the same accuracy in the
parametric approximation \cite{YB03,Y02,YBJ02} and in the microscopic
quantum dynamics approach \cite{KB02,KGB03}. Some qualitative results 
have been also presented in \cite{GGR01}. A modified version of the 
MF theory \cite{JM02} differs from the HFB method by neglecting the 
normal density, which
actually reduces the problem to a two-body one (see \cite{KH02,K02}),
removing quantum many-body effects. An attempt to describe the JILA
 $^{85}$Rb
experiments using this theory in \cite{MSJ02} has therefore met 
with little success.

The present work uses the parametric approximation \cite{YB03} in
order to describe the JILA experiments \cite{C00}. The incorporation 
of the quantum effects leads to a much better agreement with the
experimental results compared to MF theories. The results have been 
preliminarily reported in \cite{Y03}.
A more recent work \cite{KG03}, performed independently, uses the microscopic 
quantum-field-theoretic approach of 
\cite{KGB03}, in order to obtain the same goal. 
The approach of \cite{KG03} takes account of the spatial inhomogeneity 
of the condensate, and therefore yields a better agreement with the
experimental data, compared to the one presented here.

The approach of \cite{KGB03} used by \cite{KG03}, 
based on the 
numerical solution of a nonlinear partial-integro-differential 
equation, may provide the ultimate word concerning accuracy. 
But it lacks the transparency of the parametric approximation as used 
here. At the cost of avoiding inhomogeneities, the present method 
provides a greater qualitative insight, such as the occupation 
of the various non-condensate states, from which useful information can 
be derived regarding the profile of the entangled atom pair production. 
Also one can easily trace causal effects, 
such as the relation between the the excessive 
condensate loss and the effects of quantum Bose enhancement 
on the curve-crossing process (see \cite{YBJ02}).

Following the generalized parametric approximation \cite{YB03},
let us consider a system of coupled atomic and molecular fields
described by annihilation operators in the momentum representation
$\hat{\Psi }_{a}\left( {\bf p},t\right) $ and $\hat{\Psi }_{m}\left(
 {\bf p},t\right) $, respectively. The coupling of the atomic and
molecular fields contains a product of two atomic creation operators
and therefore describes the formation of entangled atomic pairs.
Spatial inhomogeneity due to the trapping potential and the effects 
of elastic collisions are neglected here.

Let the initial state of the atomic field at $t=t_{0}$  be a coherent
state of zero kinetic energy
\begin{equation}
\hat{\Psi }_{a}\left( {\bf p},t_{0}\right) |\text{in}\rangle =\left(
 2\pi \right) ^{3/2}\varphi _{a}\left( t_{0}\right) \delta \left(
 {\bf p}\right) |\text{in}\rangle  ,
\end{equation}
where $|\varphi _{a}\left( t_{0}\right) |^{2}=n_{0}\left( t_{0}\right
) $ is the initial atomic condensate density
and $|$in$\rangle $ is the  time-independent state vector in the
 Heisenberg
representation. A pair of condensate atoms forms a molecule of zero
kinetic energy. Therefore the resonant molecules can be represented by
a mean field $\varphi _{m}\left( t\right) $ as
\begin{equation}
\langle \text{in}|\hat{\Psi }_{m}\left( {\bf p},t\right)
|\text{in}\rangle =\left( 2\pi \right) ^{3/2}\varphi _{m}\left(
 t\right) \delta \left( {\bf p}\right)  ,
\end{equation}
where $|\varphi _{m}\left( t\right) |^{2}=n_{m}\left( t\right) $ is
 the molecular condensate density.
Fluctuations of the molecular field due to collisions involving 
non-condensate atoms are neglected.

The outcome of atom-molecule and molecule-molecule deactivating
collisions is introduced by adding molecular ``dump'' states. The
elimination of these states in a second-quantized description leads to
the equation of motion for the atomic field \footnote{A system of
units in which Planck's constant is $\hbar =1$ is used below.},
\begin{equation}
\fl i\dot{\hat{\Psi }}_{a}\left( {\bf p},t\right)  =\left\lbrack {p{
 } ^{2}\over 2m} + \epsilon _{a}\left( t\right) -i\gamma |\varphi
 _{m}\left( t\right) |^{2}\right\rbrack \hat{\Psi }_{a}\left( {\bf
 p},t\right) +2g^{*}\varphi _{m}\left( t\right) \hat{\Psi }^{\dag
 }_{a}\left( -{\bf p},t\right)  +i \hat{F}\left( {\bf p},t\right)  ,
 \label{Psia}
\end{equation}
where $m$ is the atomic mass, $\epsilon _{a}\left( t\right) =-{1\over
 2}\mu \left( B\left( t\right) -B_{0}\right) $ is the time-dependent 
Zeeman shift of the atom in an external magnetic field
 $B\left( t\right) $,
relative to half the energy of the molecular state, $\mu $ is the
difference in magnetic momenta of an atomic pair and a molecule, and
$B_{0}$  is the resonance value of $B$. The coupling $g$ of the
 atomic and the
molecular fields is related to the phenomenological resonance strength
$\Delta $ as $|g|^{2}=2\pi |a_{a}\mu |\Delta /m$ \cite{YBJW99}, where
 $a_{a}$  is the background elastic
scattering length. The quantum noise source $\hat{F}\left( {\bf
 p},t\right) $ and the parameter $\gamma $
describe the effect of deactivating collisions. The deactivation plays
important role in general case \cite{YB03} and is included in the
present calculations. However for the conditions of experiments
\cite{C00} the results of calculations are insensitive to these
processes (a variation of the deactivation rate coefficients from 0 to
$10^{-9}$  cm$^{3}/$s change the results by less than 1\%). 
The contribution of deactivating collisions is therefore neglected in 
following analysis. The atomic field operator can then be written in 
the form
\begin{equation}
\hat{\Psi }_{a}\left( {\bf p},t\right) =\hat{\Psi }_{a}\left( {\bf
 p},t_{0}\right) \psi _{c}\left( p,t\right) +\hat{\Psi }^{\dag
 }_{a}\left( {\bf p},t_{0}\right) \psi _{s}\left( p,t\right)  .
\end{equation}
The $c$-number functions $\psi _{c,s}\left( p,t\right) $ are
 solutions of the equations
\begin{equation}
i\dot{\psi }_{c,s}\left( p,t\right) =\left\lbrack {p{ } ^{2}\over 2m}
 + \epsilon _{a}\left( t\right) \right\rbrack  \psi _{c,s}\left(
 p,t\right) +2g^{*}\varphi _{m}\left( t\right) \psi ^{*}_{s,c}\left(
 p,t\right)  , \label{Psics}
\end{equation}
given the initial conditions $\psi _{c}\left( p,t_{0}\right) =1$,
 $\psi _{s}\left( p,t_{0}\right) =0$, and the constraint
\begin{equation}
|\psi _{c}\left( p,t\right) |^{2}-|\psi _{s}\left( p,t\right) |^{2}=1
 . \label{normpsi}
\end{equation}

The two-atom correlation functions are expressed in terms of
these solutions as
\begin{eqnarray}
\fl \langle \text{in}|\hat{\Psi }^{\dag }_{a}\left( {\bf p},t\right)
 \hat{\Psi }_{a}\left( {\bf p}^\prime ,t\right) |\text{in}\rangle
 =\left( 2\pi \right) ^{3}n_{0}\left( t\right) \delta \left( {\bf
 p}\right) \delta \left( {\bf p}^\prime \right) +n_{s}\left(
 p,t\right) \delta \left( {\bf p}-{\bf p}^\prime \right)
 \label{ncorr}
\\
\fl \langle \text{in}|\hat{\Psi }_{a}\left( {\bf p},t\right)
 \hat{\Psi }_{a}\left( {\bf p}^\prime ,t\right) |\text{in}\rangle
 =\left( 2\pi \right) ^{3}m_{0}\left( t\right) \delta \left( {\bf
 p}\right) \delta \left( {\bf p}^\prime \right) +m_{s}\left(
 p,t\right) \delta \left( {\bf p}+{\bf p}^\prime \right)
 \label{acorr} ,
\end{eqnarray}
where
\begin{equation}
n_{0}\left( t\right) =|\varphi _{a}\left( t\right) |^{2} \label{n0}
\end{equation}
is the condensate density,
\begin{equation}
\varphi _{a}\left( t\right) =\langle \text{in}|\hat{\Psi }_{a}\left(
 0,t\right) |\text{in}\rangle =\psi _{c}\left( 0,t\right) \varphi
 _{a}\left( t_{0}\right) +\psi _{s}\left( 0,t\right) \varphi ^{
*}_{a}\left( t_{0}\right)  \label{phia}
\end{equation}
is the atomic mean field,
\begin{equation}
n_{s}\left( p,t\right) =|\psi _{s}\left( p,t\right) |^{2} \label{ns}
\end{equation}
is the momentum spectrum of the non-condensate atoms, and
\begin{equation}
m_{0}\left( t\right) =\varphi ^{2}_{a}\left( t\right)  ,\qquad
 m_{s}\left( p,t\right) =\psi _{s}\left( p,t\right) \psi _{c}\left(
 p,t\right)  \label{Anden}
\end{equation}
are the anomalous densities of the condensate and non-condensate
atoms, respectively. The atomic density
\begin{equation}
\fl n_{a}\left( t\right) =\left( 2\pi \right) ^{-3}\int
 d^{3}p_{1}d^{3}p_{2}\exp\left\lbrack i\left( {\bf p}_{2}-{\bf
 p}_{1}\right) \cdot {\bf r}\right\rbrack \langle \text{in}|\hat{\Psi
 }^{\dag }_{a}\left( {\bf p}_{1},t\right) \hat{\Psi }_{a}\left( {\bf
 p}_{2},t\right) |\text{in}\rangle  ,
\end{equation}
then appears to be ${\bf r}$-independent, and comprises the sum
$n_{a}\left( t\right) =n_{0}\left( t\right) +\int d^{3}p n_{s}\left(
 p,t\right) $ of the densities of condensate atoms and of
non-condensate (entangled) atoms in a wide spectrum of kinetic
energies.

The equation of motion for the molecular mean field $\varphi
 _{m}\left( t\right) $, 
obtained by the removal of the effects of deactivating collisions, 
followed by a mean-field averaging, is
\begin{equation}
i \dot{\varphi }_{m}\left( t\right) =g m_{0}\left( t\right) +{g\over
 2\pi { } ^{2}}\int\limits^{p{ } _{c}}_{0}p^{2}d p m_{s}\left(
 p,t\right) , \label{Phim}
\end{equation}
where the cutoff $p_{c}$ in momentum space is necessary in order to
avoid divergences. The final results, after a renormalization of the
detuning $\epsilon _{a}\left( t\right) $ in a manner similar to the 
one used in \cite{HPW01}, are
insensitive to the value of the cutoff momentum if it is large
enough, so that  $p^{2}_{c}/\left( 2m\right) >\max\left( -\epsilon
 _{a}\left( t\right) \right) $.

Equations (\ref{Psics}), (\ref{normpsi}), (\ref{phia}),
(\ref{ns}), and (\ref{Anden}) lead to the following equations of 
motion for the atomic mean field, and the normal and anomalous 
densities of the non-condensate atoms,
\begin{eqnarray}
i \dot{\varphi }_{a}\left( t\right) =\epsilon _{a}\left( t\right)
 \varphi _{a}\left( t\right) +2g^{*}\varphi ^{*}_{a}\left( t\right)
 \varphi _{m}\left( t\right)
\\
i \dot{n}_{s}\left( p,t\right) =2g^{*}\varphi _{m}\left( t\right) m^{
*}_{s}\left( p,t\right) -2g\varphi ^{*}_{m}\left( t\right) m_{s}\left
( p,t\right)
\\
i \dot{m}_{s}\left( p,t\right) =\left\lbrack {p{ } ^{2}\over m} +
 2\epsilon _{a}\left( t\right) \right\rbrack m_{s}\left( p,t\right)
+2g^{*}\varphi _{m}\left( t\right) \left\lbrack 1+2n_{s}\left(
 p,t\right) \right\rbrack .
\end{eqnarray}
These equations, in combination with (\ref{Phim}) are the
momentum representation of the HFB equations used in \cite{HPW01}.
Therefore the present approach becomes mathematically equivalent to
the HFB one having, however, some advantages in simplifying the 
numerical calculations.

The results presented here were obtained by a numerical solution of 
Eqs.\ (\ref{Psics}) on a grid of values of $p$, combined with Eq.\
(\ref{Phim}). The values of $\Delta =11$ G, $B_{0}=154.9$ G, and $
|a_{a}|=450$ (in
atomic units) are taken from \cite{D02}, and the value of $\mu =
-2.23$ (in
Bohr magnetons) is taken from \cite{KH02}. The dynamics is calculated
for a linear variation of the magnetic field $B=B_{0}-\dot{B}t$ from
 162.3 G to
132 G with various sweep rates $\dot{B}$, chosen in accordance with 
the experiments \cite{C00}.

The results are sensitive to the initial value of the magnetic
field. This sensitivity can be attributed to the large crossing
width $\Gamma _{cr}$  \cite{YBJW00}, which substantially exceeds the 
range of the detuning
$\epsilon _{a}\left( t\right) $ along the full sweep. Therefore, in
 some sense, the system is
always within the resonance. As the initial values of $\varphi _{m}$,
 $\psi _{c}$, and $\psi _{s}$ were substituted the steady-state 
values calculated using the initial magnetic field as a constant. The
calculations were performed for a homogeneous BEC with the initial
density $10^{12}$  cm$^{-3}$.

\begin{figure}
\includegraphics[height=1.7in]{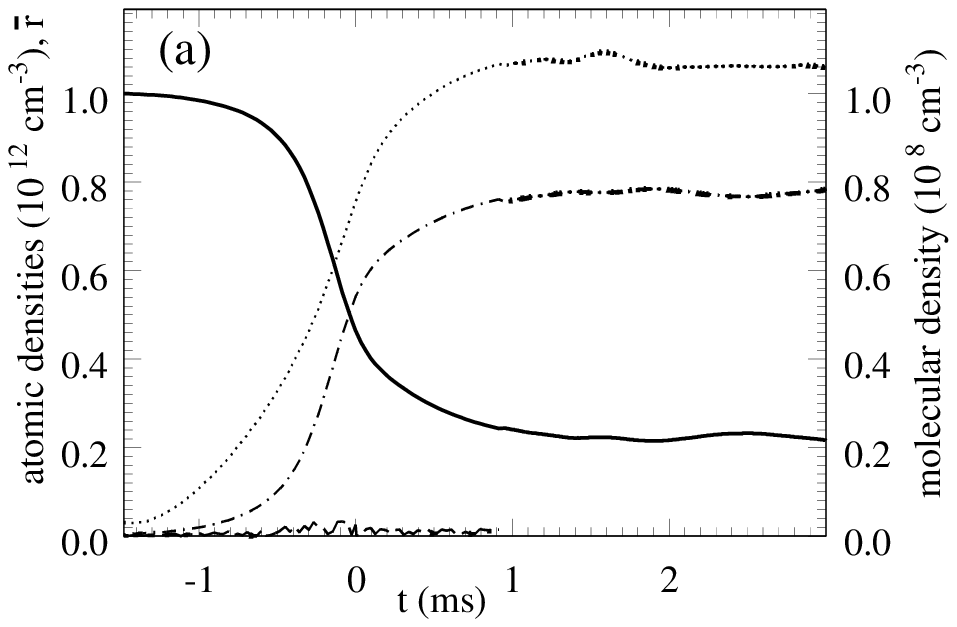}
\includegraphics[height=1.7in]{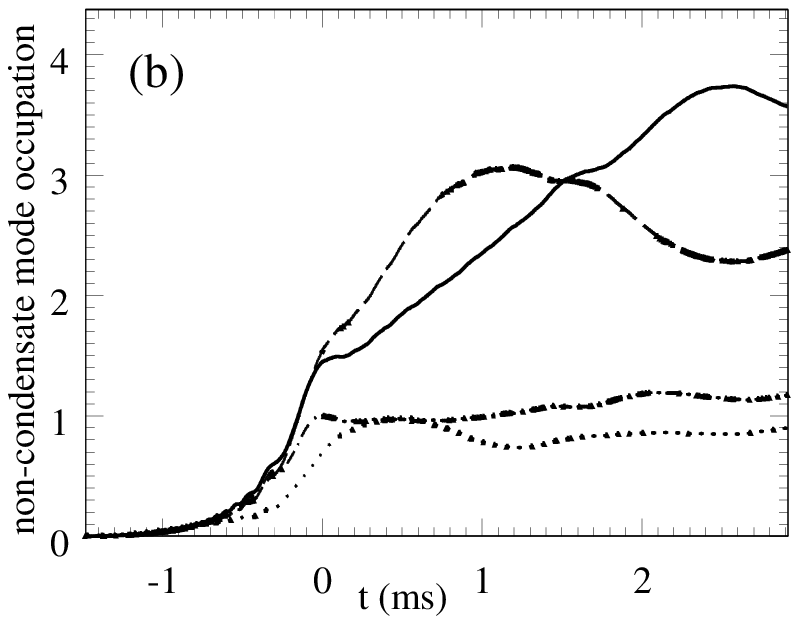}

\caption{(a) Time dependence of the densities of the atomic condensate
(\full), the molecular condensate (\broken), non-condensate atoms
(\chain) and of the mean squeezing parameter $\bar{r}$ (\protect\ref{re})
(\dotted), 
calculated for the ramp speed $\dot{B}=5$ G/ms. 
(b) The occupation
$n_{s}\left( p,t\right) $ of several non-condensate modes with
 energies $p^{2}/\left( 2m\right) = 1.1$ nK
(\chain), 4.4 nK (\full), 9.8 nK (\broken), and 17.5 nK (\dotted).}
\label{td}

\end{figure}

An example of the resulting dynamics is presented in figure \ref{td}. 
It demonstrates that the main mechanism of the condensate loss is a
conversion to non-condensate atoms by the crossing mechanism, while
the molecular occupation, and hence the deactivation losses, are
negligibly small. The effect of Bose 
enhancement on the crossing from the molecular state to 
non-condensate modes is 
proportional to the mode occupation presented in figure \ref{td}b. 

\begin{figure}
\includegraphics[height=1.7in]{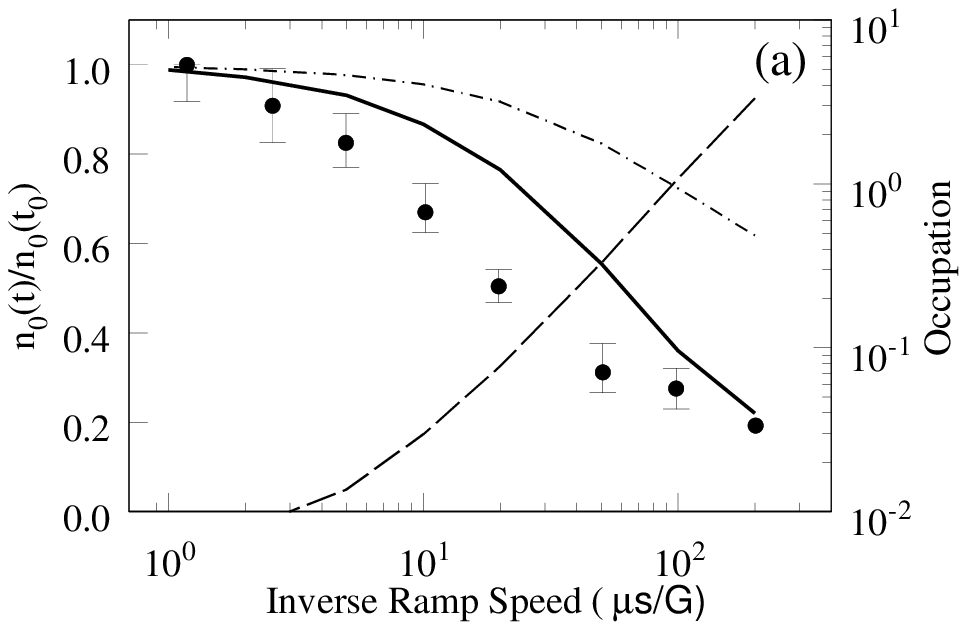}
\includegraphics[height=1.7in]{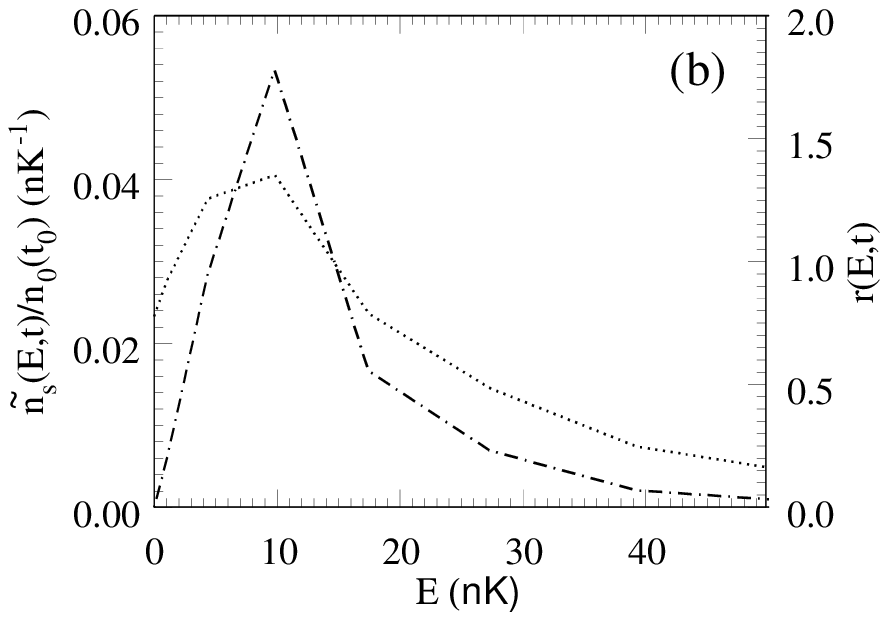}

\caption{(a) Ratio of the surviving atomic condensate density $n_{0}\left(
 t\right) $ to
the initial one $n_{0}\left( t_{0}\right) $  (\full), and the maximal 
non-condensate state
occupation (\broken), as functions of the inverse ramp speed, calculated
using the generalized parametric approximation, in comparison with the
experimental data by Cornish {\it et al.} \protect\cite{C00}
(\fullcircle), and the results of mean-field calculations based on 
\protect\cite{YBJW00} (\chain).
(b) Energy spectra of non-condensate atoms $\tilde{n}_{s}\left( E,t\right)$
(\protect\ref{ne}) (\chain) and the squeezing parameter $r(E,t)$ (\protect\ref{re}) 
(\dotted), calculated for the ramp speed $\dot{B}=5$ G/ms at the time $t=1$ ms.
} \label{loss}

\end{figure}

Figure \ref{loss}a compares the results of calculations to the
experimental data \cite{C00}. The results of mean-field
calculations using the method \cite{YBJW00} are
presented as well. It demonstrates the better agreement with the 
experimental data attained by the parametric approximation, compared 
to the MF results. The drastic increase of loss at the slower sweeps is
actually related to the effect of Bose enhancement, as demonstrated by 
the plot for the maximal value of the non-condensate state
occupation $n_{s}$  in figure \ref{loss}a. The disagreement between the
parametric approximation and the experimental data at intermediate
sweep rates is related to the effects of spatial inhomogeneity neglected
in both theories, but taken account of in the calculations \cite{KG03}.

The non-condensate atoms are formed in two-mode squeezed states,
which are entangled \cite{YB03}. Entangled atoms can have many applications
in quantum measurements, calculations, and communications (see \cite{YB03}
and references therein). The amount of entanglement can be measured
by the energy-dependent squeezing parameter $r(E,t)$ and a mean squeezing 
parameter $\bar{r}(t)$ \cite{YB03},
\begin{equation}
\fl
r(E,t) ={1\over 4}\ln{1+2n_{s}\left( p,t\right) +2|m_{s}\left( p,t\right)
|\over |1+2n_{s}\left( p,t\right) -2|m_{s}\left( p,t\right) ||}, \quad
\bar{r}\left( t\right) ={\int d^{3}p n_{s}(p,t)r({p^2\over 2 m},t)\over 
\int d^{3}p n_{s}(p,t)}
 \label{re}
\end{equation}
where $E=p^2/(2m)$ is a kinetic energy of non-condensate atoms.
In our calculations
the maximal squeezing takes place at the slowest sweep. Even in this case
the mean squeezing does not exceed the value of 1.1 (see figure \ref{td}a),
and the energy-dependent squeezing does not exceed the value of 1.4
(see figure \ref{loss}b). These values are substantially less than
$\max\bar{r}( t)\approx3.1$ and $\max r(E,t)\approx3.6$ calculated
for the weak resonance ($\Delta\approx9.5$ mG) in Na \cite{YB03}. 
This result justifies the conclusion of \cite{YB03} that a weak resonance
is preferable for formation of entangled atoms. The energy spectrum
of non-condensate atoms 
\begin{equation}
\tilde{n}_{s}\left( E,t\right) =m p n_{s}(p,t)/(2\pi^{2}),
 \label{ne}
\end{equation}
presented in figure \ref{loss}b, is rather narrow, just as in the 
case of Na \cite{YB03}.

In conclusion, this letter presents an application of the parametric
approximation \cite{YB03} to a description of BEC losses in the JILA
$^{85}$Rb experiments \cite{C00}. The method, while disregarding 
inhomogeneities, offers a rather transparent demonstration of the 
relation of the excessive losses observed to the quantum many-body 
effects of Bose enhancement, and provides detailed information on the 
nature of the entangled atom pairs produced in the process.

\end{document}